\documentclass{elsart}
\usepackage{epsfig}
\usepackage{amsmath}
\usepackage{color}
\usepackage{amssymb}
\usepackage{graphicx}
\usepackage{subfigure}

\newcommand \be{\begin{equation}}
\newcommand \ba{\begin{eqnarray}}
\newcommand \ee{\end{equation}}
\newcommand \ea{\end{eqnarray}}

\begin{document}
\runauthor{Zhou and Sornette} \markboth{A}{B}
\begin{frontmatter}
\title{Non-parametric Determination of
Real-Time Lag Structure between Two Time Series: the
``Optimal Thermal Causal Path'' Method}
\author[iggp,ess,nice]{\small{Didier Sornette}\thanksref{EM}},
\author[iggp,ecust]{\small{Wei-Xing Zhou}}
\address[iggp]{Institute of Geophysics and Planetary Physics,
University of California, Los Angeles, CA 90095}
\address[ess]{Department of Earth and Space Sciences, University
of California, Los Angeles, CA 90095}
\address[nice]{Laboratoire de Physique de la Mati\`ere Condens\'ee,
CNRS UMR 6622 and Universit\'e de Nice-Sophia Antipolis, 06108
Nice Cedex 2, France}
\address[ecust]{State Key Laboratory of
Chemical Reaction Engineering, East China University of Science
and Technology, Shanghai 200237, China}
\thanks[EM]{Corresponding author. Department of Earth and Space
Sciences and Institute of Geophysics and Planetary Physics,
University of California, Los Angeles, CA 90095-1567, USA. Tel:
+1-310-825-2863; Fax: +1-310-206-3051. {\it E-mail address:}\/
sornette@moho.ess.ucla.edu (D. Sornette)\\
http://www.ess.ucla.edu/faculty/sornette/}
\begin{abstract}

We introduce a novel non-parametric methodology to test for the
dynamical time evolution of the lag-lead structure between two
arbitrary time series. The method consists in constructing a
distance matrix based on the matching of all sample data pairs
between the two time series. Then, the lag-lead structure is
searched as the optimal path in the distance matrix landscape that
minimizes the total mismatch between the two time series, and that obeys
a one-to-one causal matching condition.
To make the solution robust
to the presence of large noise that may lead to
spurious structures in the distance matrix landscape, we then
generalize this optimal
search by introducing a fuzzy search
by sampling over all possible paths, each path being
weighted according to a multinomial logit or equivalently Boltzmann factor
proportional to the exponential of the global mismatch of this
path. We present the efficient transfer matrix method that solves
the problem and test it on simple synthetic examples to
demonstrate its properties and usefulness compared with the
standard running-time cross-correlation method. We then apply our
`Optimal Thermal Causal Path'' method to the question of the
causality between the US stock market and the treasury bond yields
and confirm our earlier results on a causal arrow of the stock
markets preceding the Federal Reserve Funds adjustments as well as
the yield rates at short maturities in the period 2000-2003.
Our application of this technique to inflation, inflation change, GDP
growth rate
and unemployment rate unearths
non-trivial ``causal'' relationships: the GDP changes lead inflation
especially since the 1980s, inflation changes leads GDP only in the
1980 decade,
and inflation leads unemployment rates since the 1970s.
In addition, our approach seems to detect multiple competing causality paths
in which one can have inflation leading GDP with a certain lag time and GDP
feeding back/leading inflation with another lag time.

\end{abstract}
%
%\begin{keyword}
%Econophysics; keyword2 \PACS 89.65.Gh; 5.45.Df
%\end{keyword}

\end{frontmatter}

\typeout{SET RUN AUTHOR to \@runauthor}

\section{Introduction}
\label{s1:intro}

Determining the arrow of causality between two time series $X(t)$ and
$Y(t)$ has a long history, especially in economics, econometrics and
finance, as it is often asked which economic variable might influence
other economic phenomena [{\it Chamberlain}, 1982; {\it Geweke}, 1984].
This question is raised in particular for the relationships between
respectively inflation and GDP, inflation and growth rate, interest rate
and stock market returns, exchange rate and stock prices, bond yields
and stock prices, returns and volatility [{\it Chan et al.}, 2001],
advertising and consumption and so on. One simple naive measure is the
lagged cross-correlation function $C_{X,Y}(\tau)=\langle X(t) Y(t+\tau)
\rangle / \sqrt{{\rm Var}[X] {\rm Var}[Y]}$, where the brackets $\langle
x \rangle$ denotes the statistical expectation of the random variable
$x$. Then, a maximum of $C_{X,Y}(\tau)$ at some non-zero positive time
lag $\tau$ implies that the knowledge of $X$ at time $t$ gives some
information on the future realization of $Y$ at the later time $t+\tau$.
However, such correlations do not imply necessarily causality in a
strict sense as a correlation may be mediated by a common source
influencing the two time series at different times. The concept of
Granger causality bypasses this problem by taking a pragmatic approach
based on predictability: if the knowledge of $X(t)$ and of its past
values improves the prediction of $Y(t+\tau)$ for some $\tau >0$, then
it is said that $X$ Granger causes $Y$ [{\it Ashley et al.}, 1980; {\it
Geweke}, 1984] (see [{\it Chen et al.}, 2004] for a recent extension to
nonlinear time series). Such a definition does not address the
fundamental philosophical and epistemological question of the real
causality links between $X$ and $Y$ but has been found useful in
practice. Our approach is similar in that it does not address the
question of the existence and tests of a genuine causality but attempts
to detect a dependence structure between two time series at non-zero
lags. We thus use the term ``causality'' in a loose sense embodying the
notion of a dependence between two time series with a non-zero lag time.

However, most economic and financial time series are not strictly stationary
and the lagged correlation and/or causality between two time series may be
changing as a function time, for instance reflecting regime switches
and/or changing
agent expectations. It is thus important to define tests of causality or of
lagged dependence which are sufficiently reactive to such regime
switches, allowing to
follow almost in real time the evolving structure of the causality.
Cross-correlation
methods and Granger causality tests require rather substantial amount of data
in order to obtain reliable conclusions. In addition,
cross-correlation techniques
are fundamentally linear measures of dependence and may miss
important nonlinear
dependence properties. Granger causality tests are most often formulated
using linear parametric auto-regressive models. The new technique introduced
in this paper, called the
``Optimal thermal causal path,'' is both non-parametric and
sufficiently general so as to
detect a priori arbitrary nonlinear dependence structures. Moreover, it is
specifically conceived so as to adapt to the time evolution of the
causality structure.
The ``Optimal thermal causal path'' can be viewed as an extension of the ``time
distance'' measure which amounts to comparing trend lines upon horizontal
differences of two time series [{\it Granger and Jeon}, 1997].

The organization of the paper is as follows. Section 2 defines
the ``Optimal thermal causal path'' method. Section 3 applies
it to simple auto-regressive models, in a first test of its
properties and limitations.
Section 4 presents an application of the Optimal thermal causal path method on
two important economic problems: the causal relationship between
the US treasury bond yields and the stock market in the aftermath of the
Internel bubble collapse and between inflation, inflation change,
gross domestic product
rate and unemployment rate in the
United States. Section 5 concludes. The Appendix presents the mathematical
algorithm underlying the construction of the Optimal thermal causal path.

\section{Definition of the ``optimal thermal causal path'' method}
\label{s1:DP}

The key ideas behind the optimal thermal causal path method can be
summarized as follows:
\begin{enumerate}
\item A distance matrix is formed which allows one to compare
systematically all values of the first time series $X(t_1)$ along
the time axis with all the values of the second time series
$Y(t_2)$, via the introduction of a distance $d(X(t_1),Y(t_2))$.
\item The causal relationship between the two time series is
searched in the form of a one-to-one mapping $t_2=\phi(t_1)$ between the
times $\{t_1\}$ of the first time series and the times $\{t_2\}$
of the second time series such that the two time series are the
closest in a certain sense, i.e., $X(t_1)$ and $Y(\phi(t_1))$
match best. We impose in addition a kind of smoothness requirement,
equivalent in most cases to continuity and monotonicity of the mapping $\phi$.
But, our ``optimal thermal causal path'' method allows
to detect situations in which the lag can jump and behave in
an arbitrary way as a function of time, as in the example
(\ref{Eq:Jump}) below.
\item The
optimal matching in step 2 is performed by introducing a weighted
average over many potential mappings in order to remove as much as
possible the influence of non-informative noises in both time
series. There is an exact mapping of this problem to a well-known
problem in statistical physics known as the directed polymer in a
quenched random potential landscape at non-zero temperature, hence
the name ``optimal thermal causal path.'' \item The resulting
mapping defines the lag between the two time series as a function
of time that best synchronizes or matches them. This thus allows
us to obtain the time evolution of the causal relationship between
the two time series.
\end{enumerate}

We now describe in details how to implement these ideas.

\subsection{Distance matrix}
\label{s2:DisMat}

To simplify, we consider time series updated in discrete
time, in units of some elementary discretization step,
taken unity without loss of generality.
Let us denote $\{X(t_1):t_1=0,...,N_1-1\}$ and
$\{Y(t_2):t_2=0,...N_2-1\}$ the two time series that we would like
to test for causality.  Note that the lengths
$N_1$ and $N_2$ of the two series can in principle be different as
our method generalizes straightforwardly to this case, but for the
sake of pedagogy, we restrict here to the case $N_1=N_2=N$. These
time series $\{X(t_1)\}$ and $\{Y(t_2)\}$ can be very different in
nature with largely different units and meanings. To make them
comparable, we normalize them by their respective standard
deviations, so that both normalized time series have comparable
typical values. From now on, the two time series $\{X(t_1)\}$ and
$\{Y(t_2)\}$ denote these normalized time series.

We introduce a distance matrix
$E_{X,Y}$ between $X$ to $Y$ with elements defined as
\begin{equation}
\epsilon(t_1,t_2) = | X(t_1)-Y(t_2) |~.
\label{Eq:Rxy}
\end{equation}
The value $| X(t_1)-Y(t_2) |$ defines the distance between the
realization of the first time series at time $t_1$ and the
realization of the second time series at time $t_2$. Other
distances can be considered and our method described below applies
without modifications for any possible choice of distances.
Depending on the nature of the time series, it may be interesting
to use others distances, which for instance put more weight on
large discrepancies $| X(t_1)-Y(t_2) |$ such as by using distances
of the form $| X(t_1)-Y(t_2) |^q$ with $q > 1$. In the following,
we do not explore this possibility and only use (\ref{Eq:Rxy}).

When $Y(t)$ is the same time series as $X(t)$, a matrix deduced from
(\ref{Eq:Rxy}) by introducing a threshold so that entries of the matrix
(\ref{Eq:Rxy}) smaller (respectively larger) than the threshold are
set to $0$ (respectively $1$) has been introduced
under the name ``recurrence
plot'' to analyze complex chaotic time series [{\it Eckmann et
al.}, 1987]. In the physical literature, the binary matrix
deduced from (\ref{Eq:Rxy}) with the use of a threshold
for two different time series is called a cross-recurrence plot.
This matrix and several of its statistical properties have been
used to characterize the cross-correlation structure between pairs
of time series [{\it Strozzia et al.}, 2002; {\it Quian Quiroga et
al.}, 2002; {\it Marwan and Kurths}, 2002; {\it Marwan et al.},
2002].

Consider the simple example in which $Y(t)=X(t-k)$ with $k>0$ fixed.
Then, $\epsilon(t_1,t_2) =0$ for $t_2=t_1+k$ and is typically
non-zero otherwise.
The detection of this causal relationship then amounts in this case to
find the line  with zero
values which is parallel to the main diagonal of the distance matrix. This line
defines the affine mapping $t_2=\phi(t_1) = t_1+k$, corresponding to
a constant translation. More generally,
we would like to determine a sequence of elements of this distance
matrix along which the elements are the smallest, as we describe next.

\subsection{Optimal path at ``zero temperature''}
\label{s2:ZeroT}

When the relationship between $X(t_1)$ and $Y(t_2)$ is more
complex than a simple constant lead-lag of the form $Y(t)=X(t-k)$,
the determination of the correspondence between the two time
series is less obvious. A first approach would correspond to
associate to each entry $X(t_1)$ of the first time series the
value $Y(t_2)$ of the second time series which makes the distance
(\ref{Eq:Rxy}) minimum over all possible $t_2$ for a fixed $t_1$.
This defines the mapping $t_1 \to t_2=\phi(t_1)$ from the
$t_1$-variable to the $t_2$-variable as
\be
\phi(t_1) = {\rm
Min}_{t_2} | X(t_1)-Y(t_2) | ~.
\label{mmbled}
\ee
Note that this
procedure analyzes each time $t_1$ independently of the others.
The problem with this approach is that it produces mappings
$t_2=\phi(t_1)$ with two undesirable properties: (i) numerous large jumps
and (ii) absence of one-to-one matching ($\phi$ is no more
a function since the curve can have overhangs and ``cliffs'') which
can also be viewed as a
backward (non-causal) time propagation. Property (i)
means that, in the presence of noise in two time series, with
large probability, there will be quite a few values of $t_1$ such that
$\phi(t_1+1)-\phi(t_1)$ is large and of the order of the total
duration $N$ of the time series. Most of the time, we
can expect lags to be slowly varying function of time and
large jumps in the function $\phi$ are not reasonable.
The second property
means that, with large probability, a given $t_1$ could be associated
with several $t_2$, and therefore there will be pairs of
times $t_1<t_1'$ such that $\phi(t_1) > \phi(t_1')$: an occurrence
in the future in the first time series is associated with an event
in the past in the second time series. This is not excluded
as lags between two time series can shift from positive to negative
as a function of time, as in our example (\ref{Eq:Jump}) below.
But such occurrences should be relatively rare in real time series
which are not dominated by noise.
Obviously, these two properties disqualify the
method (\ref{mmbled}) as a suitable construction of a time
correspondence between the two time series. This reflects the fact
that the obtained description of the lag structure between the two
time series is erratic, noisy and unreliable.

To address these two problems, we first search for a smooth mapping
$t_1 \to t_2=\phi(t_1)$:
\be
0 \leq \phi(t_1+1)-\phi(t_1) \leq 1~.
\label{jhkjhk}
\ee
In the continuous time limit, this amounts to imposing that the
mapping $\phi$ should be continuous.
Then, the correspondence $t_1 \to t_2=\phi(t_1)$ can be
interpreted as a reasonable time-lag or time-lead structure of the
two time series. For some applications, it may be desirable to
constraint even further by ensuring the differentiability (and not
only the continuity) of the mapping (in the continuous limit).
This can be done by a generalization of the global optimization
problem (\ref{mmblesddsd}) defined below by adding a path
``curvature'' energy term.  Here, we do not pursue this idea
further. Then, the causal relationship between the two time series
is searched in the form of a mapping $t_2=\phi(t_1)$ between the
times $\{t_1\}$ of the first time series and the times $\{t_2\}$
of the second time series such that the two times series are the
closest in a certain sense, i.e., $X(t_1)$ and $Y(\phi(t_1))$
match best, in the presence of these two constraints.

To implement these ideas, our first proposal is to replace the
mapping (\ref{mmbled}) determined by a local minimization by a
mapping obtained by the following global minimization: \be {\rm
Min}_{\{\phi(t_1),~t_1=0, 2, ..., N-1\}}~ \sum_{t_1=0}^{N-1} |
X(t_1)-Y(\phi(t_1)) | ~, \label{mmblesddsd} \ee under the
constraint (\ref{jhkjhk}). Note that, without the constraint
(\ref{jhkjhk}), the solution for the mapping of the minimization
(\ref{mmblesddsd}) would recover the mapping obtained from the
local minimization (\ref{mmbled}), as the minimum of the
unconstrained sum is equal to the sum of the minima. In contrast,
the presence of the continuity constraint changes
the problem into a global optimization problem.

This problem has actually a long history and has been extensively
studied, in particular in statistical physics (see [{\it Halpin-Healy
and Zhang}, 1995] for a review and references therein),
under the name of the ``random directed polymer
at zero temperature.'' Indeed, the distance matrix $E_{X,Y}$ given
by (\ref{Eq:Rxy}) can be interpreted as an energy landscape in the
plane $(t_1,t_2)$ in which the local distance $\epsilon(t_1,t_2)$
is the energy associated with the node $(t_1,t_2)$. The continuity
constraint means that the mapping defines a path or line or
``polymer'' of equation $(t_1, t_2=\phi(t_1))$ with a ``surface
tension'' preventing discontinuities. The condition that
$\phi(t_1)$ is non-decreasing translates in the fact that the
polymer should be directed (it does not turn backward and there
are no overhangs). The global minimization problem
(\ref{mmblesddsd}) translates into searching for the polymer
configuration with minimum energy. In the case where the two time
series are random, the distance matrix (and thus energy landscape)
is random, and the optimal path is then called a random directed
polymer at zero temperature (this last adjective
``at zero temperature'' will become clear
in the next section \ref{s2:FiniteT}). Of course, we are
interested in non-random time series, or at least in time series
with some non-random components: this amounts to having the
distance matrix and the energy landscape to have hopefully
coherent structures (i.e., non-white noise) that we can detect.
Intuitively, the lag-lead
structure of the two time series will reveal itself through the
organization and structure of the optimal path.

\begin{figure}[htb]
\centering
\includegraphics[width=12cm]{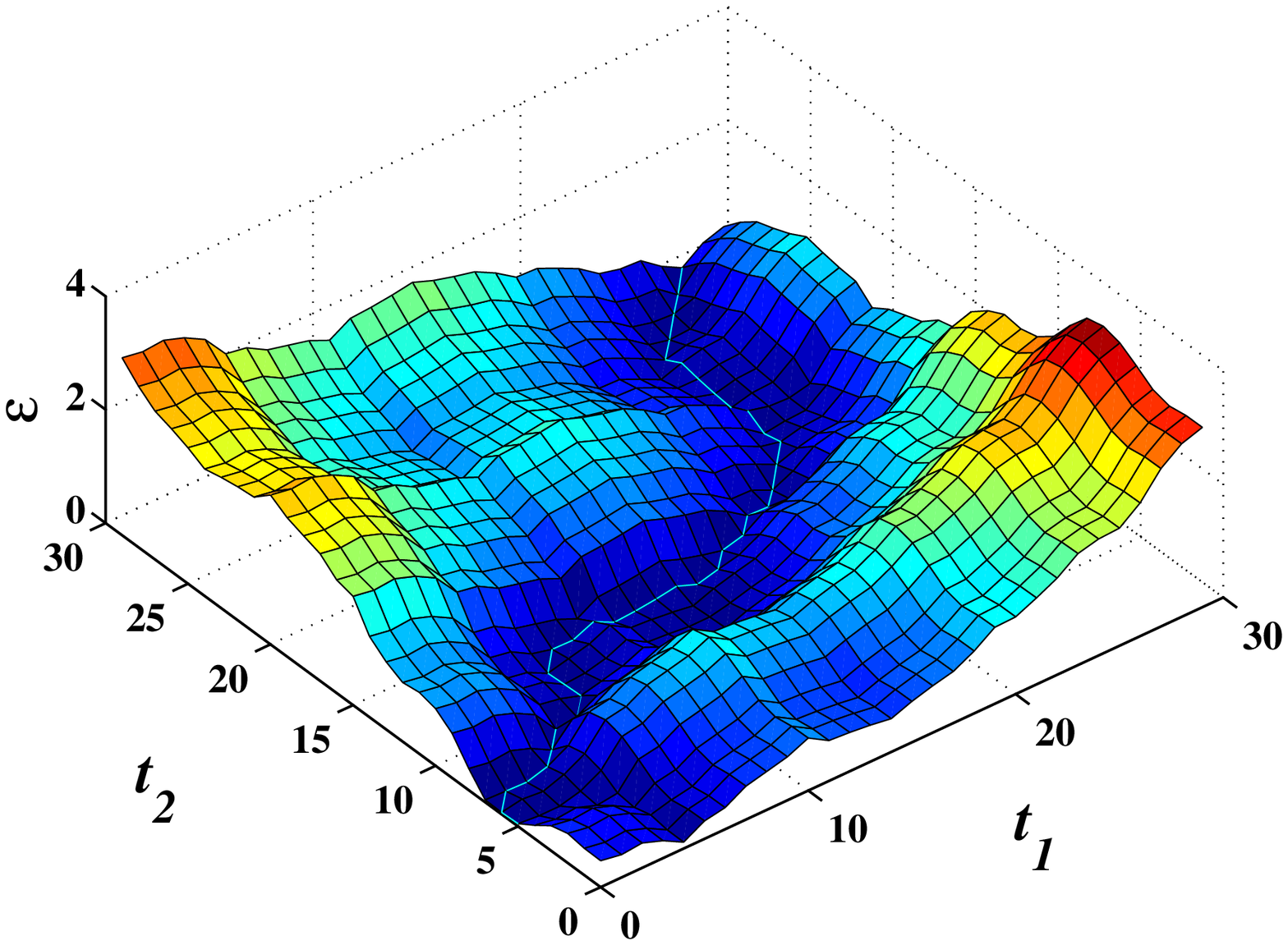}
\caption{An example of energy landscape  $E_{X,Y}$ given
by (\ref{Eq:Rxy}) for two noisy time
series and the corresponding optimal path wandering at the bottom of the
valley similarly to a river. This optimal path defines the mapping
$t_1 \to t_2 = \phi(t_1)$. }
\label{Fig:Landscape}
\end{figure}

It is important to stress the non-local nature of the optimization
problem (\ref{mmblesddsd}), as the best path from an origin to an
end point requires the knowledge of the distance matrix (energy
landscape) $E_{X,Y}$ both to the left as well as to the right of
any point in the plane $(t_1,t_2)$. There is a general and
powerful method to solve this problem in polynomial time using the
transfer matrix method [{\it Derrida et al.}, 1978; {\it Derrida
and Vannimenus}, 1983]. Figure \ref{Fig:Landscape} shows a
realization of the distance (or energy) landscape $E_{X,Y}$ given
by (\ref{Eq:Rxy}) and the corresponding optimal path.

\begin{figure}[htb]
\centering
\includegraphics[width=12cm]{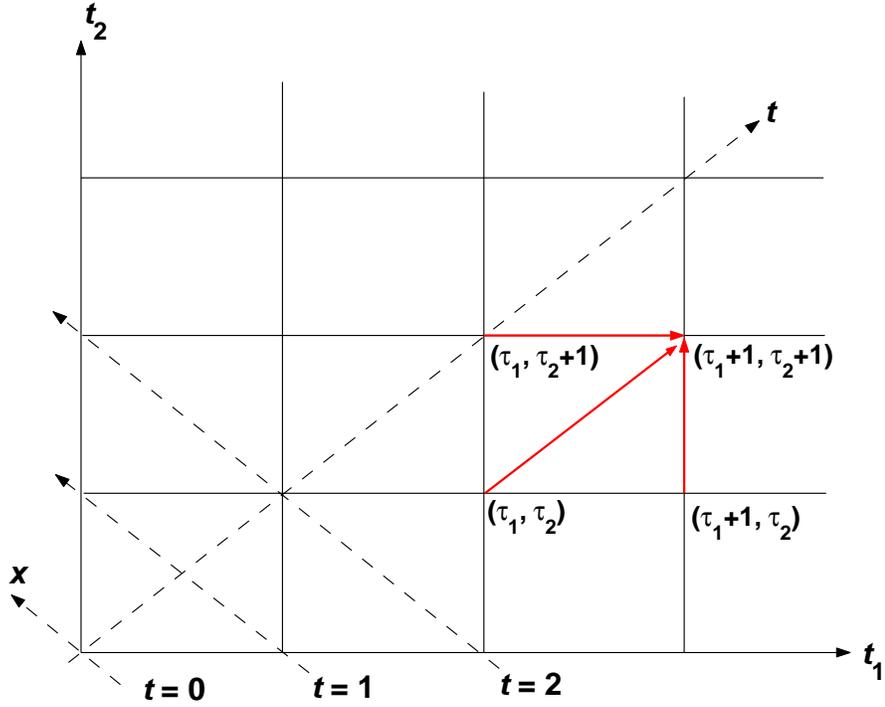}
\caption{Representation of the lattice $(t_1,t_2)$ and of the
rotated frame $(t,x)$ as defined in the text and the Appendix. We
refer to the $(t_1,t_2)$ coordinate system as the
$\square$-system (square system). We refer to the $(x,t)$ coordinate
system as the
$\triangleleft$-system (triangle system). The three arrows depict the
three moves
that are allowed from any node in one step, in accordance with the
continuity and monotonicity conditions (\ref{jhkjhk}). }
\label{Fig:TMM}
\end{figure}

The transfer matrix method can be formulated as follows. Figure
\ref{Fig:TMM} shows the $(t_1,t_2)$ plane and defines the
notations. Note in particular that the optimal path for two
identical time series is the main diagonal, so deviations from the
diagonal quantify lag or lead times between the two time series.
It is thus convenient to introduce a rotated frame $(t, x)$ as
shown in Figure \ref{Fig:TMM} such that the second coordinate $x$
quantifies the deviation from the main diagonal, hence the lead or
lag time between the two time series. In general, the optimal path
is expected to wander around, above or below the main diagonal of equation
$x(t)=0$. The correspondence between the initial frame $(t_1,t_2)$
and the rotated frame $(t, x)$ is explicited in the Appendix.

The optimal path (and thus mapping) is constructed such that it can
either go horizontally by one step from $(t_1, t_2)$
to $(t_1+1, t_2)$, vertically by one step from $(t_1, t_2)$ to
$(t_1, t_2+1)$ or along the diagonal from $(t_1, t_2)$ to
$(t_1+1, t_2+1)$. The restriction to these three possibilities embodies
the continuity condition (\ref{jhkjhk}) and the one-to-one mapping (for
vertical segments the one-to-one correspondence is ensured by the
convention to map $t_1$ to the largest value $t_2$ of the segment).
A given node $(t_1,t_2)$ in the two-dimensional lattice
carries the ``potential energy'' or distance
$\epsilon(t_1,t_2)$. Let us now denote $E(t_1,t_2)$ as the
energy (cumulative distance (\ref{mmblesddsd}))
of the optimal path starting from some origin $(t_{1,0},t_{2,0})$ and
ending at $(t_1,t_2)$. The transfer matrix method is based
on the following fundamental relation:
\begin{equation}
E(t_1,t_2) = \epsilon(t_1,t_2) + {\rm Min}\left[E(t_1-1,t_2),
E(t_1,t_2-1), E(t_1-1,t_2-1)\right]~. \label{Eq:TMM}
\end{equation}
The key insight captured by this equation is that the minimum
energy path that reaches point $(t_1,t_2)$ can only come from one of the
three points $(t_1-1,t_2)$, $(t_1,t_2-1)$ and
$(t_1-1,t_2-1)$ preceding it. Then, the minimum energy path reaching
$(t_1,t_2)$
is nothing but an extension of the minimum energy path reaching one
of these three preceding points, determined from the
minimization condition (\ref{Eq:TMM}). Then, the global optimal path
is determined as follows. One needs to
consider only the sub-lattice $(t_{1,0},t_{2,0})\times(t_1,t_2)$
as the path is directed. The
determination of the optimal path now amounts to determining
the forenode of each node in the sub-lattice
$(t_{1,0},t_{2,0})\times(t_1,t_2)$.
Without loss of generality, assume that $(t_{1,0},t_{2,0})$ is the
origin $(0,0)$. Firstly, one performs a left-to-right and
bottom-to-up scanning.
The forenode of the bottom nodes $(\tau_1,0)$ is
$(\tau_1-1,0)$, where $\tau_1=1,\cdots,t_1$. Then, one determines
the forenodes of the nodes in the second-layer at $t_2 = 1$,
based on the results of the first (or bottom) layer. This
procedure is performed for $t_2=2$, then for $t_2=3$, $\cdots$, and so
on.

The global minimization procedure is fully determined once the
starting and ending points of the paths are defined. Since the
lag-leads between two time series can be anything at any time, we
allow the starting point to lie anywhere on the horizontal axis
$t_2=0$ or on the vertical axis $t_1=0$. Similarly, we allow the
ending point to lie anywhere on the horizontal axis $t_2=N-1$ or
on the vertical axis $t_1=N-1$. This allows for the fact that one
of the two time series may precede the other one. For each given
pair of starting and ending points, we obtain a minimum path (the
``optimal directed polymer'' with fixed end-points). The minimum
energy path over all possible starting and ending points is then
the solution of our global optimization problem (\ref{mmblesddsd})
under the constraint (\ref{jhkjhk}). This equation of this global
optimal path defines the mapping $t_1 \to
t_2=\phi(t_1)$ defining the causal relationship between the two
time series.

\subsection{Optimal path at finite temperature} \label{s2:FiniteT}

While appealing, the optimization program (\ref{mmblesddsd}) under
the constraint (\ref{jhkjhk}) has an important potential drawback:
it assumes that the distance matrix $E_{X,Y}$ between the time
series $X$ to $Y$ defined by (\ref{Eq:Rxy}) is made only of useful
information. But, in reality, the time series $X(t_1)$ and
$Y(t_2)$ can be expected to contain significant amount of noise or
more generally of irrelevant structures stemming from
random realizations. Then, the distance matrix
$E_{X,Y}$ contains a possibly significant amount of noise, or in
other words of irrelevant patterns. Therefore, the global optimal
path obtained from the procedure of the previous section
\ref{s2:ZeroT} is bound to be delicately sensitive in its
conformation to the specific realizations of the noises of the two
time series. Other realizations of the noises decorating the two
time series would lead to different distance matrices and thus
different optimal paths. In the case where the noises dominates,
this question amounts to investigating the sensitivity of the
optimal path with respect to changes in the distance matrix. This
problem has actually be studied extensively in the statistical
physics literature (see [{\it Halpin-Healy and Zhang}, 1995] and
references therein). It has been shown that small changes in the
distance matrix may lead to very large jumps in the optimal path,
when the distance matrix is dominated by noise. Clearly, these
statistical properties would led to spurious interpretation of
any causal relationship between the two time series. We thus need
a method which is able to distinguish between truly informative structure
and spurious patterns due to noise.

In a realistic situation, we can hope for the existence of
coherent patterns in addition to noise, so that the optimal path
can be ``trapped'' by these coherent structures in the energy
landscape. Nevertheless, the sensitivity to specific realizations
of the noise of the two time series may lead to spurious wandering
of the optimal path, that do not reflect any genuine lag-lead
structure. We thus propose a modification of the previous global
optimization problem to address this question and make the
determination of the mapping more robust and less sensitive to the
existence of noise decorating the two time series. Of course, it
is in general very difficult to separate the noise from the
genuine signal, in absence of a parametric model. The advantage of
the method that we now propose is that it does not require any a priori
knowledge of the underlying dynamics.

The idea of the ``optimal thermal causal path'' method is the
following. Building on the picture of the optimal path as being
the conformation of a polymer or of a line minimizing its energy
$E$ in a frozen energy landscape determined by the distance
matrix, we now propose to allow from ``thermal'' excitations or
fluctuations around this path, so that path configurations with
slightly larger global energies are allowed with probabilities
decreasing with their energy. We specify the probability of a
given path configuration with energy $\Delta E$ above the absolute
minimum energy path by a multivariate logit model or equivalently by a
so-called Boltzmann weight proportional to $\exp \left[-\Delta
E/T\right]$, where the ``temperature'' $T$ quantifies how much
deviations from the minimum energy are allowed. For $T \to 0$, the
probability for selecting a path configuration of incremental
energy $\Delta E$ above the absolute minimum energy path goes to
zero, so that we recover the previous optimization problem ``at
zero temperature.'' Increasing $T$ allows to sample more and more
paths around the minimum energy path. Increasing $T$ thus allows
us to wash out possible idiosyncratic dependencies of the path
conformation on the specific realizations of the noises decorating
the two time series. Of course, for too large temperatures, the
energy landscape or distance matrix becomes irrelevant and one
looses all information in the lag-lead relationship between the
two time series. There is thus a compromise as usual between not
extracting too much from the spurious noise (not too small $T$)
and washing out too much the relevant signal (too high $T$).
Increasing $T$ allows one to obtain an average ``optimal thermal
path'' over a larger and larger number of path conformations,
leading to more robust estimates of the lag-lead structure between
the two time series. The optimal thermal path for a given $T$ is
determined by a compromise between low energy (associated with
paths with high Boltzmann probability weight) and large density
(large number of contributing paths of similar energies as larger
energies are sampled). This density of paths contributing to the
definition of the optimal thermal path can be interpreted as an
entropic contribution added to the pure energy contribution of the
optimization problem of the previous section \ref{s2:ZeroT}. In a
sense, the averaging over the thermally selected path
configurations provides an effective way of averaging over the
noise realizations of the two time series, without actually having
to resampling the two times series. This intuition is
confirmed by our tests below which show that the signal-over-noise ratio
is indeed increased significantly by this ``thermal'' procedure.

Let us now describe how we implement this idea. It is convenient
to use the rotated frame $(t,x)$ as defined in Figure \ref{Fig:TMM},
in which $t$ gives the coordinate along the main diagonal of the
$(t_1,t_2)$ lattice and $x$ gives the coordinate in the
transverse direction from the main diagonal. Of course, the origin
$(t_1=0,t_2=0)$
corresponds to $(x=0,t=0)$. Note that the constraint that the path is
directed allows us to interpret $t$ as an effective time and $x$ as
the position
of a path at that ``time'' $t$.
Then, the optimal thermal path
trajectory $\langle{x(t)}\rangle$ is obtained by the following formula
\begin{equation}
      \langle{x(t)}\rangle = \sum_x {x
G_\triangleleft(x,t)/G_\triangleleft(t)}~.
      \label{Eq:MeanX}
\end{equation}
In this expression, $G_\triangleleft(x,t)$ is the sum of Boltzmann
factors over all paths $\mathcal{C}$ emanating from $(0,0)$ and
ending at $(x,t)$ and $G_\triangleleft(t) = \sum_x
G_\triangleleft(x,t)$. In statistical physics,
$G_\triangleleft(x,t)$ is called the partition function constrained
to $x$ while $G_\triangleleft(t)$ is the total partition function
at $t$. Then, $G_\triangleleft(x,t)/G_\triangleleft(t)$ is nothing
but the probability for a path be at $x$ at ``time'' $t$. Thus,
expression (\ref{Eq:MeanX}) indeed defines $\langle{x}\rangle$ as
the (thermal) average position at time $t$. It is standard to call
it ``thermal average'' because $G$ is made of the Boltzmann
factors that weight each path configuration. The intuition is to
imagine the polymer/path as fluctuating randomly due to random
``thermal kicks'' in the quenched random energy landscape. In the
limit where the temperature $T$ goes to zero,
$G_\triangleleft(x,t)/G_\triangleleft(t)$ becomes the Dirac
function $\delta[x-x_{DP}(t)]$ where $x_{DP}(t)$ is the position
of the global optimal path determined previously in section
\ref{s2:ZeroT}. Thus, for $T \to 0$, expression (\ref{Eq:MeanX})
leads to $\langle{x}\rangle = x_{DP}(t)$, showing that this
thermal procedure generalizes the previous global optimization
method. For non-vanishing $T$, the optimal thermal average
$\langle{x(t)}\rangle$ given by (\ref{Eq:MeanX}) takes into
account the set of the neighboring (in energy) paths which allows
one to average out the noise contribution to the distance matrix.
The Appendix gives the recursion relation that allows us to
determine $G_\triangleleft(x,t)$. This recursion relation uses the
same principle and has thus the same structure as expression
(\ref{Eq:TMM}) [{\it Wang et al.}, 2000].

Similarly to expression (\ref{Eq:MeanX}), the variance of the trajectory of the
optimal thermal path reads
\begin{equation}\label{Eq:VarX}
\sigma^2_{x} = \sum_x {\left(x-\langle{x}
\rangle\right)^2G_\triangleleft(x,t) /G_\triangleleft(t)}~.
\end{equation}
The variance $\sigma^2_{x}$ gives a measure of the uncertainty in
the determination of the thermal optimal path and thus an estimate
of the error in the lag-lead structure of the two time series as
seen from this method.

\section{Numerical tests on simple examples}
\label{s1:NumSim}

\subsection{Construction of the numerical example} \label{s2:model}

We consider two stationary time series $X(t_1)$ and $Y(t_2)$,
and construct
$Y(t_2)$ from $X(t_1)$ as follows:
\begin{equation}
Y(t_2) = a X(t_2-\tau) + \eta ~, \label{Eq:LinRegMod}
\end{equation}
where $a$ is a constant, $\tau$ is the time lag, and the noise
$\eta \sim N(0,\sigma_\eta)$ is serially uncorrelated.

The time series $X(t_1)$ itself is generated from an AR
process:
\begin{equation}
X(t_1) = b X(t_1-1) + \xi~, \label{Eq:XiAR}
\end{equation}
where $b<1$ and the noise $\xi \sim N(0,\sigma_\xi)$ is serially
uncorrelated. The factor $f=\sigma_\eta/\sigma_\xi$ quantifies the
amount of noise degrading the causal relationship between $X(t_1)$
and $Y(t_2)$. A small $f$ corresponds to a strong causal
relationship. A large $f$ implies that $Y(t_2)$ is mostly noise
and becomes unrelated to $X(t_1)$ in the limit $f \to \infty$.
Specifically, ${\rm Var}[X] = \sigma_\xi^2/\left(1-b^2\right)$ and
\be{\rm Var}[Y] = a^2 {\rm Var}[X] + \sigma_\eta^2 = \sigma_\xi^2
\left( \frac{a^2}{1 - b^2} + f^2 \right) = \sigma_\xi^2 \left(
\frac{a^2{\rm Var}[X]}{\sigma_\xi^2} + f^2 \right) ~. \ee

In our simulations, we take $\tau=5$, $a=0.8$, $b=0.7$, and
$\sigma_\xi=1$ and consider time series of duration $N=100$.

For a given $f$, we obtain the optimal zero-temperature path by
using the transfer-matrix method (\ref{Eq:TMM}) explained in
section \ref{s2:ZeroT} for 19 different starting positions around
the origin and similarly 19 different ending positions around the
upper-right corner at coordinate $(99, 99)$. This corresponds to
solve $19 \times 19$ transfer matrix optimization problems. The
absolute optimal path is then determined as the path which has the
smallest energy over all these possible starting and ending
points. We also determine the optimal thermal paths
$\langle{x(t)}\rangle$, for different temperatures, typically from
$T=1/5$ to $10$, using the relation (\ref{Eq:RecurG:t1t2}) for the
partition function and the definition (\ref{Eq:t2mean}) for the
average transverse path trajectory (given in the Appendix).

Figure \ref{FigXa} shows that transverse trajectory $x(t)$ as a
function of the coordinate $t$ along the main diagonal for
$f=1/10$ and for temperatures $T=0$, $1/5$, $1$, and $10$. This graph
corresponds to the case where we retrict our attention to
paths with fixed imposed starting (origin) and ending (coordinates ($99, 99$)
on the main diagonal) points. This restriction is relaxed as we explain
above and apply below to prevent from the boundary effects clearly
visible in Figure \ref{FigXa}. Figure
\ref{FigXb} shows the corresponding standard deviation defined by
(\ref{Eq:VarX}) of the thermal average paths.

The impact of the temperature is nicely illustrated by plotting
how the energy of an optimal thermal path depends on its
initial starting point $x(0)=x_0$ (and ending point taken with the same
value $x(99)=x(0)$). For a given $x_0$ and
temperature $T$, we determine the thermal optimal path and then
calculate its energy $e_T(x_0)$ by the formula
\begin{equation}\label{Eq:e}
      e_T(x_0) = \frac{1}{2(N-|x_0|)-1}\sum_{t=|x_0|}^{2N-1-|x_0|}
      \sum_x {\epsilon(x,t)G_\triangleleft(x,t)/G_\triangleleft(t)}~.
\end{equation}
By construction, the time lag between the two time series is
$\tau=5$ so that we should expect $e_T(x_0)$ to be minimum for
$x_0=\tau=5$. Figure \ref{Fig:SZC:SP:E} plots $e_T(x_0)$ as a
function of the average of the path $\langle x(x_0)\rangle$ with
different starting points $x_0$ for different temperatures $T$
respectively equal to $1/50$, $1/5$, $1/2$, $1$, $2$, $5$, and
$10$ and for $f=1/2$. One can observe a large quasi-degeneracy for
small temperatures, so that it is difficult to identify what is
the value of the lag between the two time series. The narrow trough at
$\langle x(x_0)\rangle = 5$ for the smallest temperatures, while
at the correct value, is not clearly better than negative values of
$\langle x(x_0)\rangle$. In contrast,
increasing the temperature produces a well-defined quadratic
minimum bottoming at the correct value $\langle x(x_0)\rangle=\tau=5$
and removes the degeneracies observed for the smallest temperatures. This
numerical experiment illustrates the key idea underlying the
introduction of the thermal averaging in section \ref{s2:FiniteT}:
too small temperatures lead to optimal paths which are exceedingly
sensitive to details of the distance matrix, these details being
controlled by the specific irrelevant realizations of the noise
$\eta$ in expression (\ref{Eq:LinRegMod}). The theoretical
underpinning of the transformation from many small competing
minima to well-defined large scale minima as the temperature
increases, as observed in Figure \ref{Fig:SZC:SP:E}, is well
understood from studies using renormalization group methods [{\it
Bouchaud et al.}, 1991].

Figure \ref{Fig:SZC:TnF} further demonstrates the role of the
temperature for different amplitudes of the noise $\eta$. It shows
the position $\overline{\langle x \rangle}$ as a function of $T$
for different relative noise level $f$. Recall that
$\langle{x(t)}\rangle$ is the optimal thermal position of the path
for a fixed coordinate $t$ along the main diagonal, as defined in
(\ref{Eq:MeanX}). The symbol $\overline{\langle x \rangle}$
expresses an additional average of ${\langle x \rangle}$ over all
the possible values of the coordinate $t$: in other words,
$\overline{\langle x \rangle}$ is the average elevation (or
translation) of the optimal thermal path above (or below) the
diagonal. This average position is an average measure
(along the time series) of the lag/lead time between the two time series,
assuming that this lag-lead time is the same for all times. In our
numerical example, we should obtain $\overline{\langle x \rangle}$
close to or equal to $\tau=5$. Figure \ref{Fig:SZC:TnF} shows the
dependence of $\overline{\langle x \rangle}$ as a function of $T$
for different values of $f$.

Obviously, with the increase of the signal-to-noise ratio of the
realizations which is proportional to $1/f$, the accuracy of the
determination of $\tau$ improves. For a noise level $f$,
$\overline{\langle x \rangle}$ approaches the correct value
$\tau=5$ with increasing $T$. The beneficial impact of the
temperature is clearer for more noisy signals (larger $f$).
It is interesting to notice that an ``optimal range'' of
temperature appears for large noise level.

\subsection{Test on the detection of jumps or change-of-regime in time lag}
\label{s2:jump}

We now present synthetic tests of the efficiency of the optimal
thermal causal path method to detect multiple changes of regime
and compare the results with a standard correlation analysis
performed in moving windows of different sizes. Consider the
following model
\begin{equation}
Y(i)=\left\{
\begin{array}{ll}
   0.8X(i) + \eta, & 1\le i \le 50\\
   0.8X(i-10) + \eta, & 51\le i \le 100\\
   0.8X(i-5) + \eta, & 101\le i \le 150\\
   0.8X(i+5) + \eta, & 151\le i \le 200\\
   0.8X(i) + \eta, & 201\le i \le 250\\
\end{array}
\right.~. \label{Eq:Jump}
\end{equation}
In the sense of definition (\ref{Eq:LinRegMod}), the time series $Y$
is lagging behind $X$ with
$\tau = 0$, $10$, $5$, $-5$ (this negative lag time corresponds to
$X(t)$ lagging behind $Y(t)$), and $0$ in five successive time
periods of $50$ time steps each. The time series $X$ is assumed to be the
first-order AR process (\ref{Eq:XiAR}) and
$\eta$ is a Gaussian white noise. Our results are essentially the same
when $X$ is itself a white Gaussian random variable. We use $f=1/5$
in the simulations presented below.

Figure \ref{Fig:Jump:C} shows the standard cross-correlation
function calculated over the whole time interval $ 1\le i \le 250$
of the two time series $X$ and $Y$ given by (\ref{Eq:Jump}), so as
to compare with our method.
Without further information, it would be difficult to conclude
more than to say that the two time series are rather
strongly correlated at zero time lag. It would
be farfetched to associate the tiny secondary peaks of the
correlation function at $\tau=\pm 5$ and $10$ to genuine lags or
lead times between the two time series. And since, the correlation
function is estimated over the whole time interval, the time
localization of possible shifts of lag/leads is impossible.

Before presenting the results of our method, it is instructive
to consider a natural extension of the correlation analysis, which
consists in estimating the correlation function in a moving window
$[i+1-D,i]$ of length $D$, where $i$ runs from $D$ to $250$.
We then estimate the lag-lead time $\tau_D(i)$ as the value
that maximizes the correlation function in each window $[i+1-D,i]$.
We have used $D=10$, $20$, $50$, and $100$ to investigate
different compromises ($D=10$ is reactive but does not give
statistically robust estimates while $D=100$ gives statistically more
robust estimates but is less reactive to abrupt changes of lag).
The local lags $\tau_D(i)$ thus obtained are shown in
Fig.~\ref{Fig:Jump:TauD} as a function of the
running time $i$. For $D=10$, this method identifies
successfully the correct time lags in the first, third, fourth,
and fifth time periods, while $\tau_D(i)$ in the second time
period is very noisy and fails to unveil the correct value $\tau = 10$.
For $D=20$, the correct time lags in the five time periods are
identified with large fluctuations at the boundaries between two
successive time periods. For $D=50$, five successive time lags are
detected but with significant delays compared to their actual
inception times, with in addition high interspersed fluctuations.
For $D=100$, the delays of the detected inception times of each period
reach about 50 time units, that is, comparable to the width of each
period, and the method fails completely for this case.

Let us now turn to our optimal thermal causal path method.
We determine the average thermal path (transverse trajectory $x(i)$
as a function of
the coordinate $i$ along the main diagonal) starting at the origin,
for four different temperatures $T=2$,
1, $1/2$, and $1/5$. Figure~\ref{Fig:Jump:TauT} plots $x(i)$ as
a function of $i$. The time lags in the five time periods are
recovered clearly. At the joint points between the successive time periods,
there are short transient crossovers from one time lag to the next.
Our new method clearly outperforms the above cross-correlation analysis.

The advantage of our new method compared with the moving
cross-correlation method for two time series with varying time
lags can be further illustrated by a test of predictability. It is
convenient to use an example with unidirectional causal lags (only positive
lags) and not with bidirectional jumps as exemplified by (\ref{Eq:Jump}).
We thus consider a case in which $X$ leads $Y$ in
general and use the following model
\begin{equation}
Y(i)=\left\{
\begin{array}{ll}
   0.8X(i) + \eta, & 1\le i \le 50\\
   0.8X(i-10) + \eta, & 51\le i \le 100\\
   0.8X(i-5) + \eta, & 101\le i \le 150\\
   0.8X(i-8) + \eta, & 151\le i \le 200\\
\end{array}
\right.~. \label{Eq:Pred}
\end{equation}
At each instant $i$ considered to be the ``present,'' we perform
a prediction of $Y(i+1)$ for ``tomorow'' at $i+1$ as follows.
We first estimate the
instantaneous lag-lead time $\tau(i)$. The first estimation uses the
running-time cross-correlation method which delivers
$\tau(i)=\tau_D(i)$. The second estimation is the average
thermal position $\tau(i)=\max\{[x(i)],0\}$ using the optimal
thermal causal path method where the operator $[\cdot]$ takes the
integral part of a number. We construct the prediction for $Y(i+1)$ as
\be
  Y(i+1) = 0.8 X(i+1-\tau(i))~.
  \ee
In this prediction set-up, we assume that we have full knowledge of the
model and the challenge is only to calibrate the lag.
The standard deviations
of the prediction errors are found for the cross-correlation
method respectively equal to 2.04 for $D=10$, 0.41 for $D=20$, and
1.00 for $D=50$. Using the optimal thermal path, we find a
standard deviation of the prediction errors of 0.45 for $T=2$,
0.39 for $T=1$, 0.33 for $T=1/2$, and 0.49 for $T=1/5$.
Our optimal causal thermal path method thus
outperforms and is much more stable than the classic cross-correlation
approach.

\section{Applications to economics}
\label{s1:Appl}

\subsection{Revisiting the causality between
the US treasury bond yield and the stock market antibubble
since August 2000} \label{s2:Slaving}

In a recent paper [{\it Zhou and Sornette}, 2004], we have found
evidence for the following causality in the time period
from October 2000 to september 2003: stock market
$\to$ Fed Reserve (Federal funds rate) $\to$
short-term yields $\to$ long-term yields (as well as a direct and
instantaneous influence of the stock market on the long-term yields).
These conclusions were based on 1) lagged cross-correlation
analysis in running windows and 2) the dependence of the parameters
of a ``log-periodic power law'' calibration to the yield time series
at different maturities (see [{\it Sornette and Johansen}, 2001;
{\it Sornette and Zhou}, 2002; {\it Sornette}, 2003] for recent
exposition of the method and synthesis of the main results on a variety
of financial markets).

Let us now revisit this question by using the optimal thermal
causal path method. The data consist in the S\&P 500 index, the
Federal funds rate (FFR), and ten treasury bond yields spanning
three years from 2000/09/09 to 2003/09/09. The optimal thermal
paths $x(i)$'s of the distance matrix between the monthly returns
of the S\&P 500 index with each of the monthly relative variations
of the eleven yields are determined for a given temperature $T$,
giving the corresponding lag-lead times $\tau(i)=x(i)$'s as a
function of present time $i$. Fig.~\ref{Fig:Yield:dT20:T4} shows
these $\tau(i)$'s for $T=1$, where positive values correspond to
the yields lagging behind or being caused by the S\&P 500 index
returns. The same analysis was performed also for $T=10$, $5$,
$2$, $1$, $1/2$ and $1/5$, yielding a very consistent picture,
confirming indeed that $\tau$ is positive for short-term yields
and not significantly different from zero for long-term yields, as
shown in Fig.~\ref{Fig:Yield:dT20:T4}. One can also note that the
lag $\tau(i)$ seems to have increased with time from September
2000 to peak in the last quarter of 2003.

We also performed the same analysis with weakly and quarterly data of the
returns and yield changes. The results (not shown) confirm the results obtained
at the monthly time scale. This analysis seems to confirm the
existence of a change of regime in the arrow of causality between
the S\&P 500 index and the Federal Funds rate: it looks as if the
Fed (as well as the short term yields) started to be influenced by
the stock market after a delay following the crash in 2000,
waiting until mid-2001 for the causality to be revealed. The
positivity of the time lag shows the causal ``slaving'' of the
yields to the stock index. This phenomenon is consistent with the
evidence previously presented in [{\it Zhou and Sornette}, 2004]
and thus provides further evidence on the causal arrow flowing
from the stock market to the treasury yields. The instantaneous
lag-lead functions $\tau(t)$ provide actually much clearer
signatures of the causality than our previous analysis: compare
for instance with the cross-correlation coefficient shown in
figure 10 of [{\it Zhou and Sornette}, 2004]. From an economic
view point, we interpret these evidences, that the FRB is causally
influenced by the stock market (at least for the studied period),
as an indication that the stock markets are considered as proxies
of the present and are conditioning the future health of the economy,
according to the FRB model of the US economy. In a related study,
causality tests performed by Lamdin [2003] also confirm that stock
market movements precede changes in yield spread between corporate
bonds and government bonds. Abdulnasser and Manuchehr [2002] have
also found that Granger causality is unidirectionally running from
stock prices to effective exchange rates in Sweden.

\subsection{Are there any causal relationship between inflation and
gross domestic product (GDP) and inflation and unemployment in the USA?}
\label{s2:inflationGDP}

The relationship between inflation and real economic output
quantified by GDP has been discussed many times in the last several
decades. Different theories have suggested that the impact of
inflation on the real economy activity could be either neutral,
negative, or positive. Based on Mundell's story that higher
inflation would lower real interest rates [{\it Mundell}, 1963],
Tobin [1965] argued that higher inflation causes a shift from money to
capital investment and raise output per capita.
On the contrary, Fischer [1974 suggested a negative
effect, stating that higher inflation resulted in a shift from
money to other assets and reduced the efficiency of transactions
in the economy due to higher search costs and lower productivity.
In the middle, Sidrauski [1967] proposed a neutral effect
where exogenous time preference fixed the long-run real interest
rate and capital intensity. These arguments are based
on the rather restrictive assumption that the Philips curve (inverse
relationship between inflation and unemployment), taken in addition
to be linear, is valid.

To evaluate which model characterizes better real economic
systems, numerous empirical efforts have been performed. Fama [1982]
applied the money demand theory and the rational
expectation quality theory of money to the study of inflation in the
USA and observed a negative relation during the post-1953 period.
Barro [1995] used data for around 100 countries from
1960 to 1990 to assess the effects of inflation on economic output
and found that an increase in average inflation led to a reduction
of the growth rate of real per capita GDP, conditioned on the fact that the
inflation was high. Fountas et al. [2002]
used a bivariate GARCH model of inflation
and output growth and found evidence that higher inflation and
more inflation uncertainty lead to lower output growth in the
Japanese economy. Apergis [2004] found that inflation
affected causally output growth using a univariate GARCH models to
a panel set for the G7 countries.

Although cross-country regressions explain that output growth
often obtains a negative effect from inflation, Ericsson et al.
[2001] argued that these results are not robust and demonstrated that
annual time series of inflation and the log-level of output for
most G-7 countries are cointegrated, thus rejecting the existence
of a long-run relation between output growth and inflation. A
causality analysis using annual data from 1944 to 1991 in Mexico
performed by Shelley and Wallace [2004] showed that it is
important to separate the changes in inflation into predictable
and unpredictable components whose differences respectively had a
significant negative and positive effect on real GDP growth. Huh
[2002] and Huh and Lee [2002] utilized a vector autoregression
(VAR) model to accommodate the potentially important departure
from linearity of the Phillips curve motivated by a strand of
theoretical and empirical evidence in the literature suggesting
nonlinearity in the output-inflation relationship. The empirical
results indicated that their model captured the nonlinear features
present in the data in Australia and Canada. This study implies
that there might exists a non-linear causality from inflation to
economic output. It is therefore natural to use our novel method
to detect possible local nonlinear causality relationship.

Our optimal thermal causal path method is applied to the GDP
quarterly growth rates paired with the inflation rate updated
every quarter on the one hand and with the quarterly changes of the inflation
rates on the other hand, for the
period from 1947 to 2003 in the USA. The GDP growth
rate, the inflation rate and and the inflation rate changes have been
normalized by
their respective standard deviations. The
inflation and inflation changes are calculated from the monthly
customer price index (CPI) obtained from  the Fed II database (federal reserve
bank). Eight different temperatures $T=50, 20, 10$, $5$, $2$,
$1$, $1/2$, and $1/5$ have been investigated.

Figure \ref{FigSZC_3data} shows the data used for the analysis, that is,
the normalized inflation rate, its normalized quarterly change and
the normalized GDP growth rate from 1947 to 2003.

Figure \ref{FigSZC_InfGDP5Tau} shows the lag-lead times
$\tau(t)=x(t)$'s (units in year) for the pair (inflation, GDP
growth) as a function of present time $t$ for $T=2$ and for 19
different starting positions (and their ending counterparts) in
the $(t_1,t_2)$ plane, where positive values of $\tau(t)=x(t)$
correspond to the GDP lagging behind or being caused by inflation.
This figure is representative of the information at all the
investigated temperatures. Overall, we find that $\tau$ is
negative in the range $-2$ years $ \leq \tau \leq 0$ year,
indicating that it is more the GPD which leads inflation than the
reverse. However, this broad-brush conclusion must be toned down
somewhat at a finer time resolution as two time periods can be
identified in figure \ref{FigSZC_InfGDP5Tau}:
\begin{itemize}
\item From 1947 (and possibly earlier) to early 1980s, one can
observe two clusters,
one with negative $-2$ years $ \leq \tau = x(t) \leq 0$ years implying
that the GDP has a positive causal effect on future inflation,
and another with positive $0$ years $ \leq \tau = x(t) \leq 4$ years implying
that inflation has a causal effect on GDP with a longer lag.
\item From the mid-1980s to the present, there is not doubt that it is
GDP which has had the dominating causal impact on future inflation lagged by
about $1-2$ years.
\end{itemize}

In summary, our analysis suggests that the interaction between GDP
and inflation is more subtle than previously discussed. Perhaps
past controversies on which one causes the other one may be due to
the fact that, to a certain degree, each causes the other with
different time lags. Any measure of a causal relationship allowing
for only one lag is bound to miss such subtle interplay. It is
interesting to find that GDP impacts on future inflation with a
relatively small delay of about one year while inflation has in
the past influenced future GDP with a longer delay of several
years.

Figure \ref{FigSZC_dInfGDP5Tau} shows the
lag-lead times $\tau(t)=x(t)$'s (units in year) for the pair
(inflation change, GDP) as a
function of present time $t$ for $T=2$ and for 19 different starting
positions (and their
ending counterparts) in the $(t_1,t_2)$ plane, where positive values
of $\tau(t)=x(t)$ correspond
to the GDP lagging behind or being caused by inflation change. Due to the
statistical fluctuations, we cannot conclude on the existence of
a significant causal relationship between inflation change and GDP, except
in the decade of the 1980s for which there is strong causal effect of
a change of inflation on GDP. The beginning of this decade was
characterized by a strong
decrease of the inflation rate from a two-digit value
in 1980, following a vigorous monetary policy implemented
under the Fed's chairman Paul Volker. The end of the 1970s and the
better half of the 1980s were
characterized by an almost stagnant GDP.
In the mid-1980s, the GDP started to grow again at a strong pace. It
is probably
this lag between the signifant reduction of inflation in the first
half of the 1980s
and the raise of the GDP growth that we detect here.  Our analysis may
help in improving our understanding in the intricate relationship
between different economic
variables and their impact on growth and on stability
and in addressing the difficult problem of model errors, that
Cogley and Sargent [2004] have argued to be the cause for the lack of
significant
action from the Fed in the 1970s.

%http://usinfo.state.gov/products/pubs/oecon/chap7.htm

Figure \ref{FigSZC_InfUnemp5Tau} shows the
lag-lead times $\tau(t)=x(t)$'s (units in year)
for the pair (inflation, unemployment rate) as a
function of present time $t$ for $T=2$ and for 19 different starting
positions (and their
ending counterparts) in the $(t_1,t_2)$ plane, where positive values
of $\tau(t)=x(t)$ correspond
to the unemployment rate lagging behind or being caused by inflation.
We use quaterly data from 1948 to 2004 obtained from the Fed II
database (federal reserve
bank). This figure
is representative of the information at all the investigated temperatures.
\begin{itemize}
\item From 1947 (and possibly earlier) to 1970, one can observe large
fluctuations
with two clusters, suggesting a complex causal relationship between
the two time series, similarly to the situation discussed above for the
(inflation, GDP) pair.
\item From 1970 to the present, there is not doubt that inflation has
predated and ``caused'' unemployment in the sense of the optimal
thermal causal path method.
It is also noteworthy that the lag between unemployment and inflation has
disappeared in recent years. From a visual examination of figure
ZZZZ, we surmise
that what is detected is probably related to the
systematic lags between inflation and employment in the four large
peak pairs: (1970 for
inflation; 1972 for employment), (1975 for inflation; 1976 for unemployment),
(1980 for inflation;
1983 for unemployment) and (1991 for inflation; 1993 for unemployment).
\end{itemize}
One standard explanation for a causal impact of inflation on unemployment
is through real wage: if inflation goes faster than the adjustment of salaries,
this implies that real wages are decreasing, which favors
employment according to standard economic theory,
thus decreasing unemployment. Here, we find that surges of inflation
``cause'' increases and not decreases of unemployments. Rather than
an inverse relationship between synchronous inflation and unemployment
(Philips curve), it seems that a better description of the data is a direct
lagged relationship, at least in the last thirty years. The combination
of increased inflation and unemployment has been known as ``stagflation''
and caused policymakers to abandon the notion of an exploitable
Phillips curve trade-off (see for instance [Lansing, 2000]).
Our analysis suggests a more complex multivariate description which
requires taking into account inflation, inflation change, GDP, unemployment
and their expectations by the agents, coupled all together through
a rather complex network of lagged relationships. We leave this for
a future work.

\section{Concluding remarks}
\label{s1:concl}

In summary, we have developed a novel method for the detection of
causality between two time series, based on the search for
a robust optimal path in a distance matrix. Our optimal
thermal causal path method determines the thermal average paths
emanating from different starting lag-lead times in the distance
matrix constructed from the two original time series and choose
the one with minimal average mismatch (``energy''). The main
advantage of our method is that it enables us to detect causality
locally and is thus particularly useful when the causal relation
is nonlinear and changes intermittently. An advantage of the method
is that it is robust with respect to noise, i.e., it does
not attribute causal relationships
between two time series from patterns in
the distance matrix that may arise randomly. This robustness
is acquired by using the ``thermal'' averaging procedure which
provides a compromise between optimizing the matching between the
two time series and maximizing the local density of optimal paths
to ensure a strong relationship.

We have applied this
method to the stock market and treasury bond yields and confirmed
our earlier results in \cite{ZS04PA} on a causal arrow of the
stock markets preceding the Federal Reserve Funds adjustments as
well as the Yield rates at short maturities. Another application to the
inflation and GDP growth rate and to unemployment have unearthened
non-trivial ``causal'' relationships: the GDP changes lead inflation
especially since the 1980s, inflation changes leads GDP only in the
1980 decade,
and inflation leads unemployment rates since the 1970s.

Our approach seems to detect multiple competing causality paths
with intertwinned arrows of causality in which
one can have inflation leading GDP with a certain lag time and GDP
feeding back/leading inflation with another lag time. This suggests that
the predictive skills of models with one-way causality are fundamentally
limited and more elaborate measurements as proposed here and models
with complex feedbacks are necessary to account for the multiple
lagged feedback mechanisms present in the economy.

\section*{Appendix A: Recursive scheme of partition function}
\label{s2:RecursiveG}

In order to calculate the thermal average position
$\langle{x(t)}\rangle$ for $t=0,1,2,\cdots$, over all possible
path in the distance matrix landscape,
one needs to determine the values at all nodes of $G_{\triangleleft}(x,t)$,
defined in equation (\ref{Eq:MeanX}) and subsequent paragraph. For clarity,
we present firstly the recursive relation
in the $(t_1,t_2)$ coordinates and then transform it into the $(x,t)$
coordinates. The transformation from the coordinates $(t_1,t_2)$
to $(x,t)$ is
\begin{equation}
\left\{
   \begin{array}{ccl}
    x &=& t_2-t_1~, \\
    t &=& t_2+t_1~.
    \end{array}
\right. \label{Eq:AxesTransform}
\end{equation}
Note that the $x$ has a different unit from $t_2$, which have a
factor of $\sqrt{2}$ geometrically.

If two time series are perfectly causally related (they are the
same up to a factor), then the optimal path is the diagonal, that
is, made of the diagonal bonds of the square lattice, or
alternatively the nodes on the diagonals. Since the ``energy''
(i.e., local mismatch defined by expression (\ref{Eq:Rxy})) is
defined only on the nodes, a path has a Boltzmann weight contributed
only by the nodes and there is no contribution from bonds. We should thus
allow path not only along the horizontal and vertical segments of
each square of the lattice but also along the main diagonal of
each square. The directedness means that a given path is not allowed
to go backward on any of the three allowed moves. As illustrated
in Figure \ref{Fig:TMM}, in order to arrive at $(t_1+1, t_2+1)$, the path
can come from $(t_1+1, t_2)$ vertically, $(t_1, t_2+1)$
horizontally, or $(t_1, t_2)$ diagonally. The recursive equation
on the Boltzmann weight factor is thus
\begin{subequations}
\begin{equation}\label{Eq:RecurG:t1t2}
      G(t_1+1,t_2+1) = [G(t_1+1,t_2) + G(t_1,t_2+1) + G(t_1,t_2)]
      e^{-\epsilon(t_1+1,t_2+1)/T}~,
\end{equation}
where $\epsilon(t_1+1,t_2+1)$ is loal energy determined by
the distance matrix element (\ref{Eq:Rxy})
at node $(t_1+1,t_2+1)$.

Using the axes transformation (\ref{Eq:AxesTransform}),
Eq.~(\ref{Eq:RecurG:t1t2}) can be rewritten in the following form
\begin{equation}\label{Eq:RecurG:xt}
      G_\triangleleft(x,t+1) = [G_\triangleleft(x-1,t) +
      G_\triangleleft(x+1,t) + G_\triangleleft(x,t-1)]
      e^{-\epsilon(x,t)/T}~.
\end{equation}
\end{subequations}

\section*{Appendix B: Relations between the two schemes}

Consider a $t$-slide in the $\triangleleft$-scheme, that is, in
the $x,t$ coordinates system. There are $t+1$ nodes on the
$t$-slide. For simplicity, we denote the $t+1$ partition functions
as $G_i$, $i=1,2,\cdots,t+1$, and denote $G= \sum_{i=1}^{t+1}G_i$.
We define two thermal averages of the transverse fluctuations for
the $t$-slide in the $\square$-scheme and the
$\triangleleft$-scheme, respectively:
\begin{subequations}
\begin{equation}\label{Eq:t2mean}
      \langle{t_2(t)}\rangle = \sum_{t_2=0}^{t} t_2G_{t_2+1} /G
\end{equation}
\begin{equation}\label{Eq:Xmean}
      \langle{x(t)}\rangle = \sum_{x=-t:2:t}
      xG_{(x+t+2)/2} / G
\end{equation}
\end{subequations}
Posing $i=(x+t+2)/2$, Eq.~(\ref{Eq:Xmean}) becomes \be
\langle{x(t)}\rangle = \sum_{i=1}^{t+1} [2i-(t+2)]G_i/G =
2\sum_{i=0}^{t} iG_i/G - t ~. \ee
   We have
\begin{equation}\label{Eq:taut1t2}
      \tau(t) \triangleq \langle{x(t)}\rangle = 2\langle{t_2(t)}\rangle -
      t~.
\end{equation}
Actually, this expression (\ref{Eq:taut1t2}) can be derived alternatively as
follows. Consider the optimal position at time $t\gg 0$. We have
$t_1=t-\langle{t_2(t)}\rangle$ and $t_2=\langle{t_2(t)}\rangle$
statistically. Using $\tau=t_2-t_1$, we reach (\ref{Eq:taut1t2}).
It is also easy to show that the standard deviation of the
position of the path is $\sigma_{\tau} =
2\sigma_{\langle{t_2}\rangle}$.

\bigskip
{\textbf{Acknowledgments:}}

We are grateful to X.-H. Wang for fruitful discussion and to N.
Marwan for the permission of the use of his MATLAB programs
(http://www.agnld.uni-potsdam.de) at the early stage of this work.

\clearpage

%FIGURE 1

\clearpage

%FIGURE 2

\clearpage
%FIGURE 3
\begin{figure}
\centering
  \subfigure[]{
     \label{FigXa} %% label for first subfigure
     \begin{minipage}[b]{0.47\textwidth}
       \centering
       \includegraphics[width=7cm]{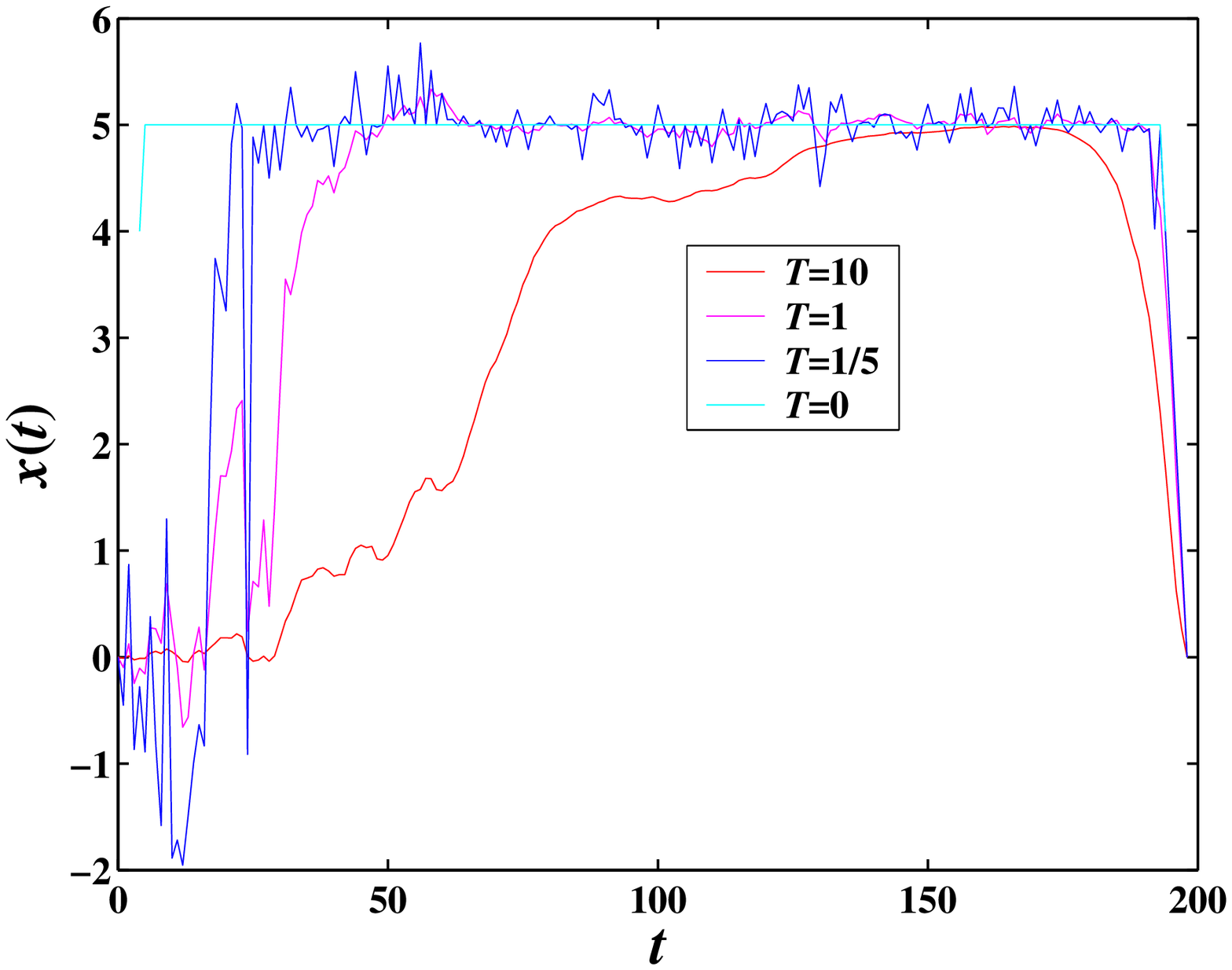}
     \end{minipage}}
     \hspace{0.1cm}
  \subfigure[]{
     \label{FigXb} %% label for first subfigure
     \begin{minipage}[b]{0.47\textwidth}
       \centering
       \includegraphics[width=7cm]{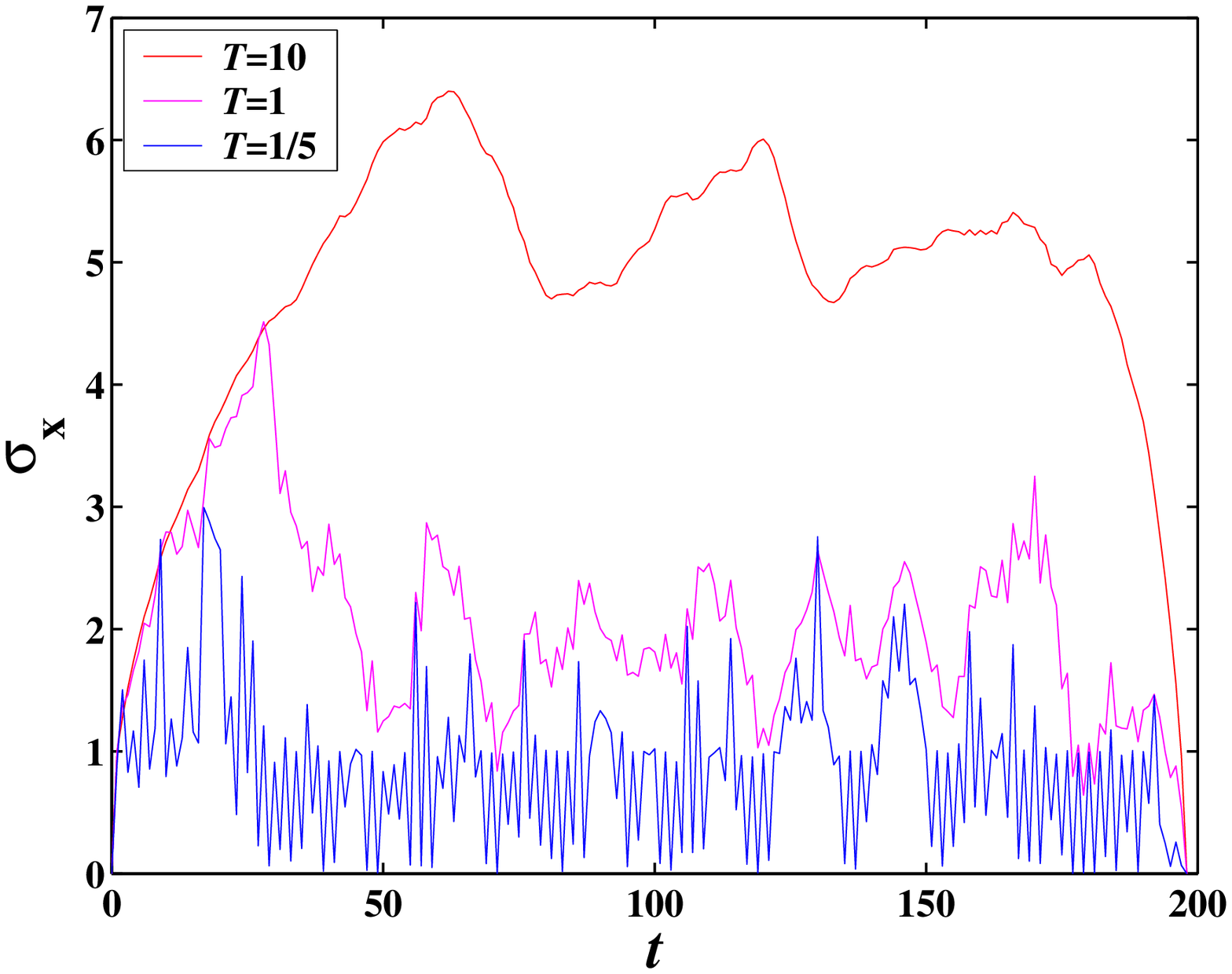}
     \end{minipage}}
  \caption{(a) Thermal average $\langle{x(t)}\rangle$ of the
transverse fluctuations with respect to
$t$ for $T=10$, $1$, and $1/5$ and the directed polymer. (b) The
uncertainty $\sigma_{x}$ of the thermal average paths for
different temperatures. All the paths are constrained to start from
the diagonal $(t_1=0, t_2=0)$ and to return to it at $(t_1=99, t_2=99)$.}
  \label{FigXab} %% label for entire figure
\end{figure}

\clearpage
%FIGURE 4
\begin{figure}[htb]
\centering
\includegraphics[width=12cm]{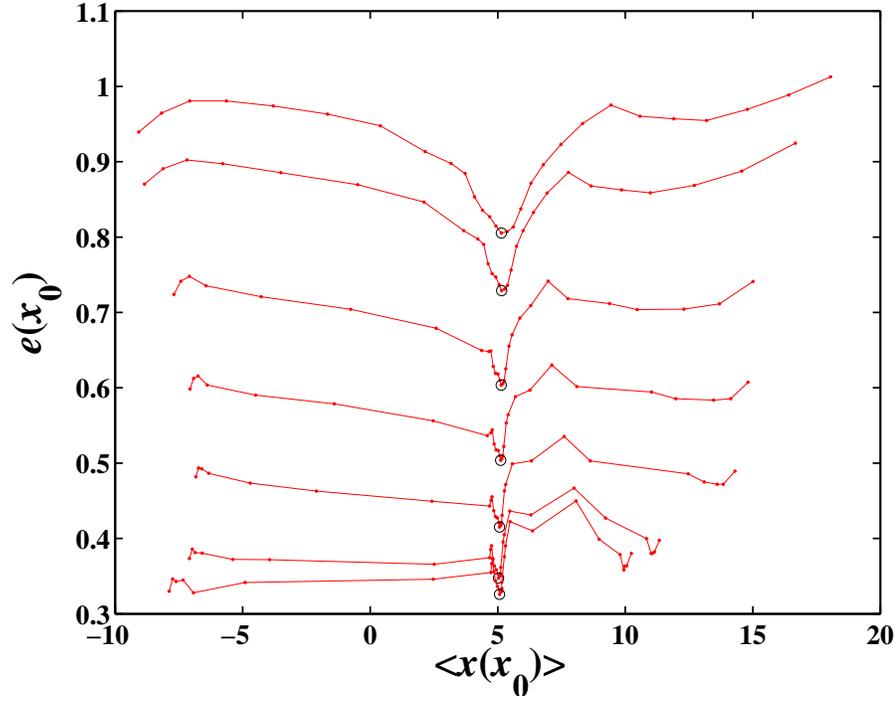}
\caption{Dependence of the thermal average energy $e_T(x_0)$ of
the optimal thermal path as a function of the average $\langle
x(x_0)\rangle$ defined in turn by the coordinate of its starting
point $(t=|x_0|, x=x_0)$ for different temperatures given by
$T=1/50$, $1/5$, $1/2$, $1$, $2$, $5$ and $10$ from bottom to top
and for $f=1/2$.} \label{Fig:SZC:SP:E}
\end{figure}

\clearpage

%FIGURE 5
\begin{figure}[htb]
\centering
\includegraphics[width=12cm]{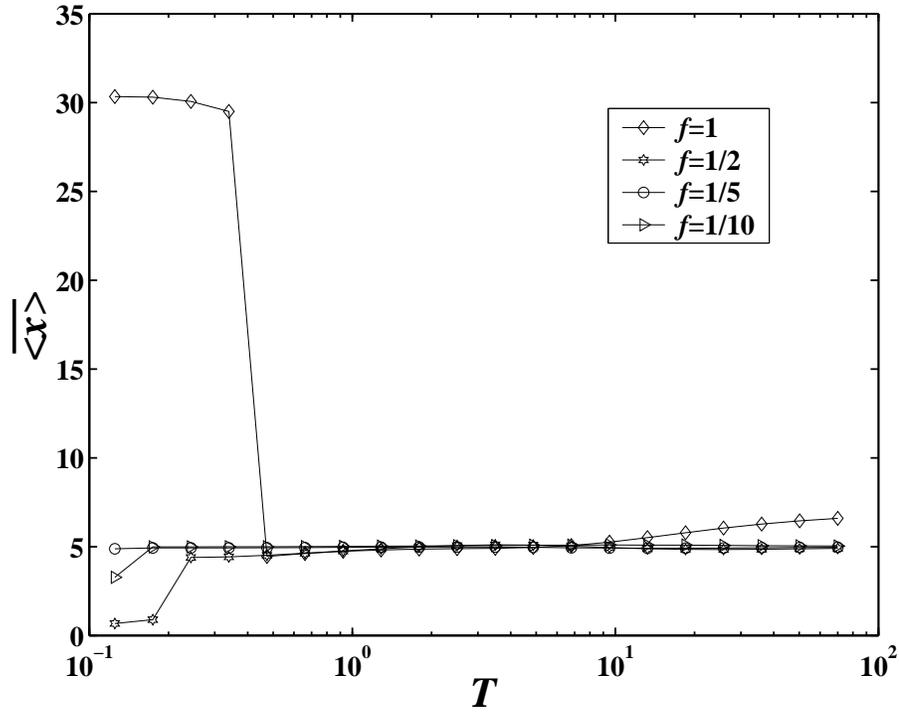}
\caption{Dependence of $\overline{\langle{x}\rangle}$ upon noise
level $f$ and temperature $T$.} \label{Fig:SZC:TnF}
\end{figure}

\clearpage

%FIGURE 6
\begin{figure}[htb]
\centering
\includegraphics[width=12cm]{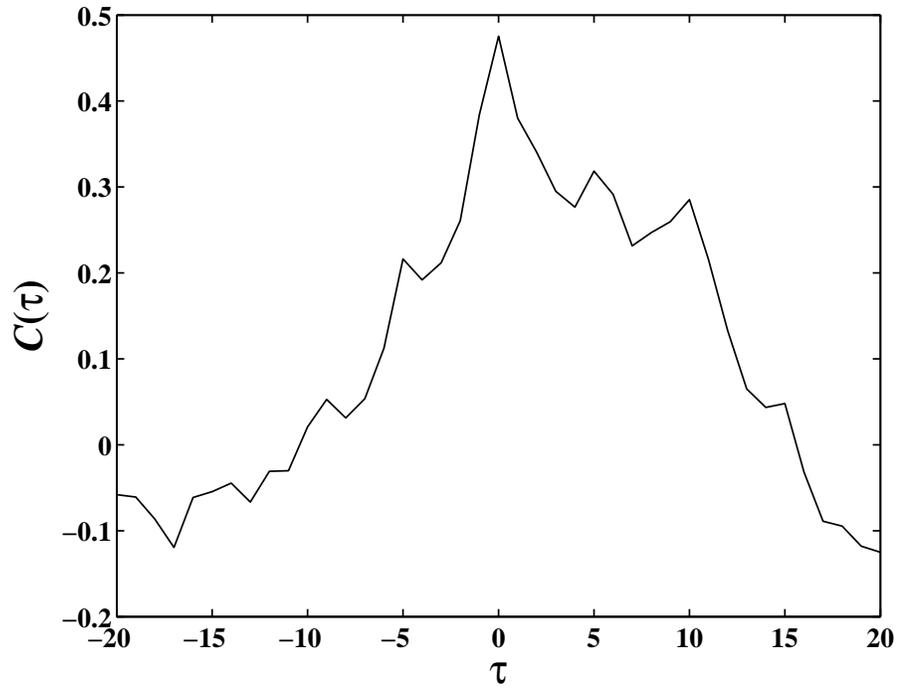}
\caption{Standard cross-correlation function of two time series
with varying time lags $\tau = 0$, $10$, $5$, $-5$, and $0$
as defined in equation (\ref{Eq:Jump}).} \label{Fig:Jump:C}
\end{figure}

\clearpage

%FIGURE 7
\begin{figure}[htb]
\centering
\includegraphics[width=12cm]{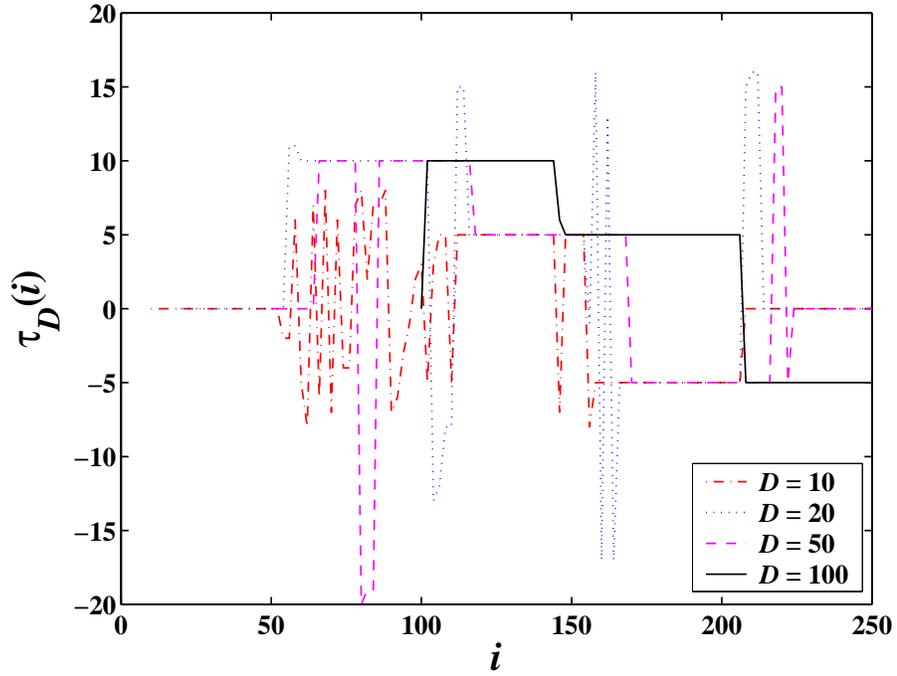}
\caption{Local cross-correlation analysis of the
two time series defined by (\ref{Eq:Jump}) with (\ref{Eq:XiAR})  using moving
windows of sizes $D=10$, $20$, $50$, and
$100$. The value $\tau_D(i)$ of the lag that makes maximum the
local cross-correlation function in each window
$[i+1-D,i]$ is plotted as a function of
the right-end time $i$. The true time lags as defined in (\ref{Eq:Jump})
are respectively $\tau = 0$, $10$, $5$, $-5$ and $0$ in five successive time
periods of $50$ time steps each.} \label{Fig:Jump:TauD}
\end{figure}

\clearpage

%FIGURE 8
\begin{figure}[htb]
\centering
\includegraphics[width=12cm]{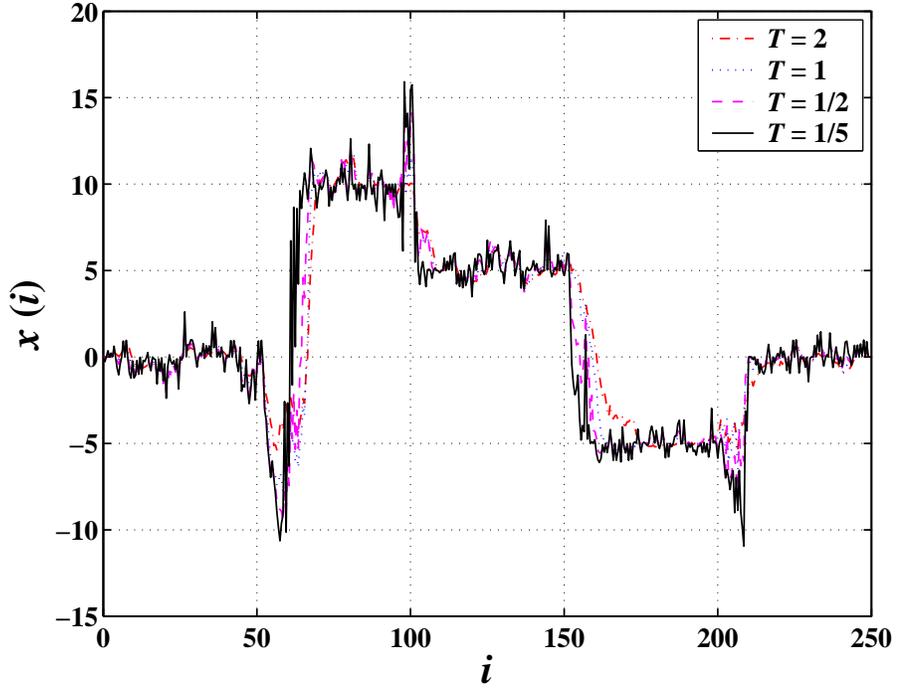}
\caption{Average thermal path (transverse trajectory $x(i)$ as a function of
the coordinate $i$ along the main diagonal) starting at the origin,
for four different temperatures ($T=2$ (dotted-dash), $T=1$
(dotted), $T=0.5$ (dashed), and $0.2$ (continuous))
obtained by applying the
optimal thermal causal path method to the synthetic time series
(\ref{Eq:Jump}) with (\ref{Eq:XiAR}).}
\label{Fig:Jump:TauT}
\end{figure}

\clearpage

%FIGURE 9
\begin{figure}[htb]
\centering
\includegraphics[width=12cm]{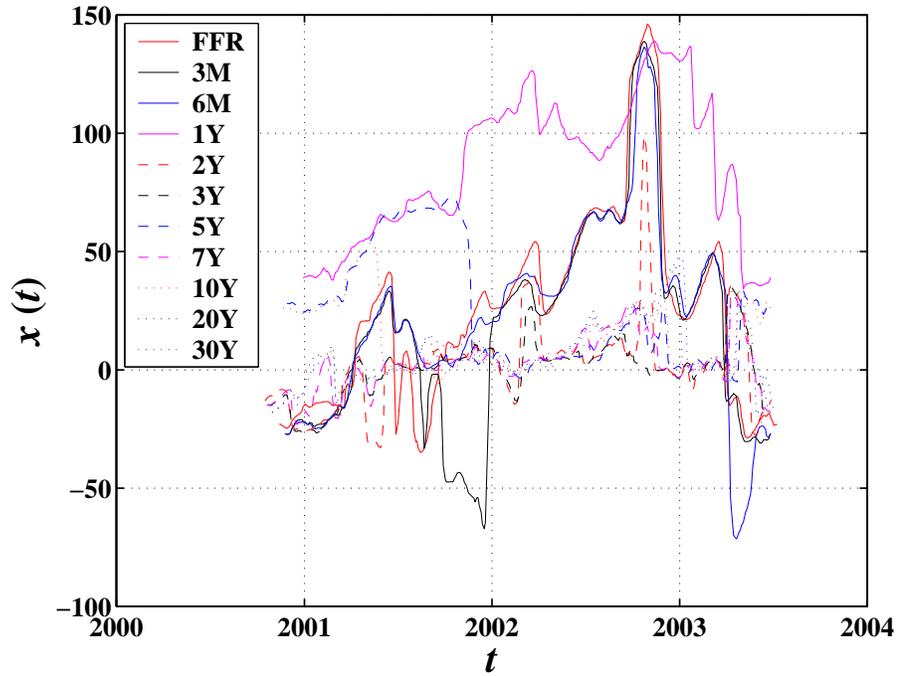}
\caption{Instantaneous lags between the S\&P 500 index and the
Federal funds rate (FFR), and between the S\&P 500 index and each
of ten treasury bond yields, calculated using the optimal thermal
causal path method at temperature $T=1$ using monthly returns for
the S\&P 500 index and monthly relative variations for the Yields.
Positive lags corresponds to the yields lagging behind the S\&P
500 index. } \label{Fig:Yield:dT20:T4}
\end{figure}

\clearpage

%FIGURE 10
\begin{figure}[htb]
\centering
\includegraphics[width=12cm]{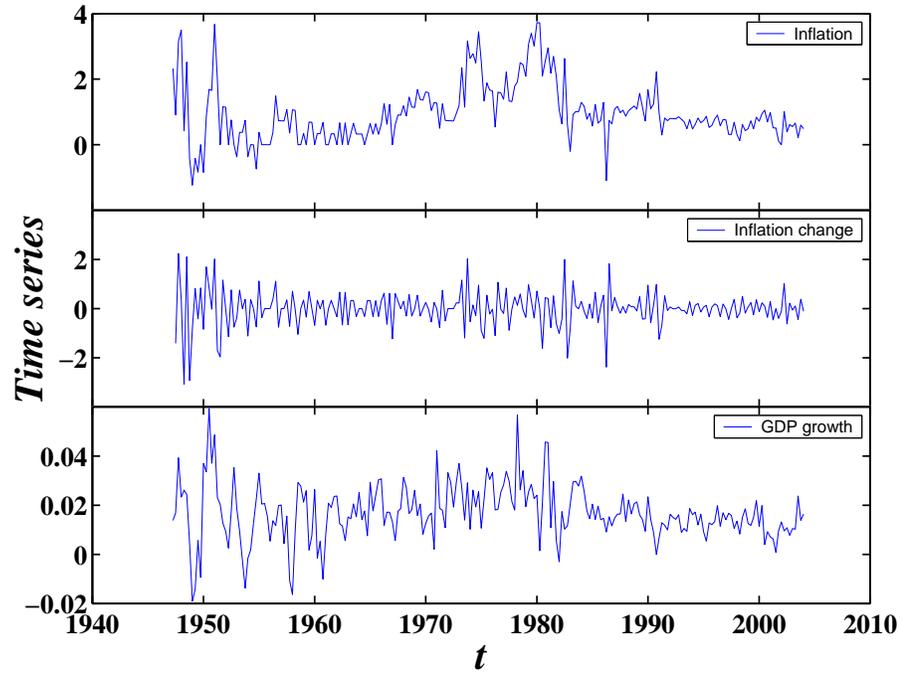}
\caption{Data used in our analysis, that is, the normalized
inflation rate, its normalized quarterly change, the normalized
GDP growth rate and the normalized unemployment rate from 1947 to
2003. } \label{FigSZC_3data}
\end{figure}

\clearpage

%FIGURE 11
\begin{figure}[htb]
\centering
\includegraphics[width=12cm]{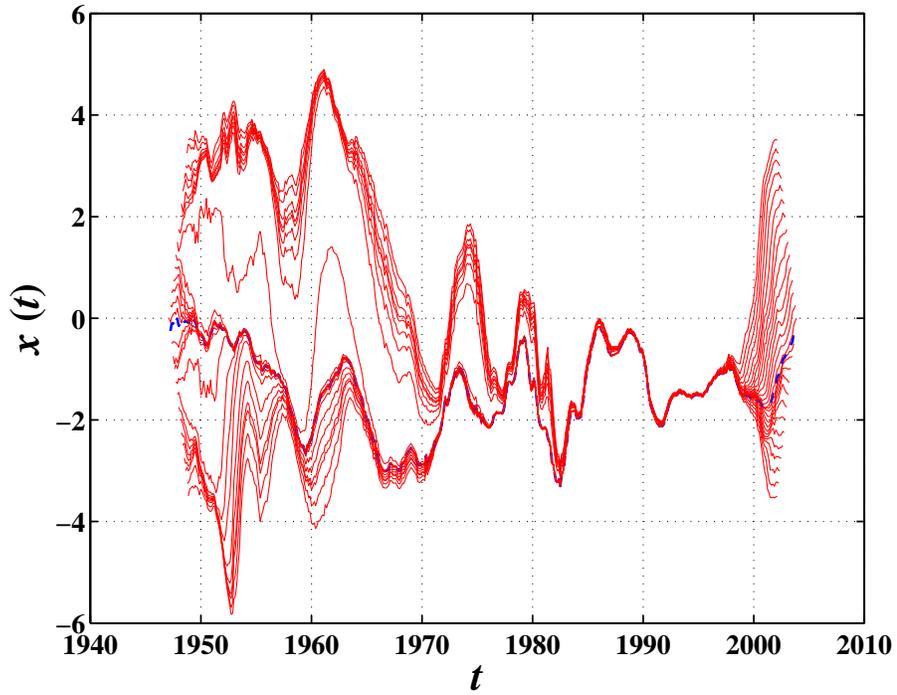}
\caption{Lag-lead times $\tau(t)=x(t)$'s (units in year) for the
pair (inflation, GDP) as a
function of present time $t$ for $T=2$ and for 19 different starting
positions (and their
ending counterparts) in the $(t_1,t_2)$ plane, where positive values
of $\tau(t)=x(t)$ correspond
to the GDP lagging behind or being caused by inflation.
The dashed blue line is the optimal path with the minimal ``energy.''
}
\label{FigSZC_InfGDP5Tau}
\end{figure}

\clearpage

%FIGURE 12
\begin{figure}[htb]
\centering
\includegraphics[width=12cm]{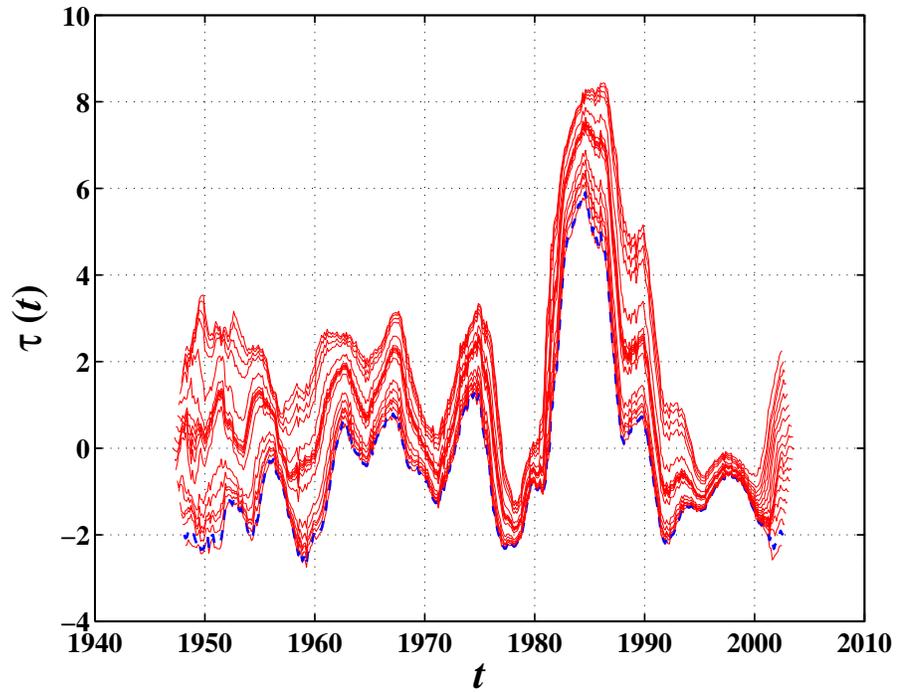}
\caption{Same as figure \ref{FigSZC_InfGDP5Tau} for the pair
(inflation change, GDP). Positive values
of $\tau(t)=x(t)$ correspond
to the GDP lagging behind or being caused by inflation change.
The dashed blue line is the optimal path with the minimal ``energy.'' }
\label{FigSZC_dInfGDP5Tau}
\end{figure}

\clearpage

%FIGURE 13
\begin{figure}[htb]
\centering
\includegraphics[width=12cm]{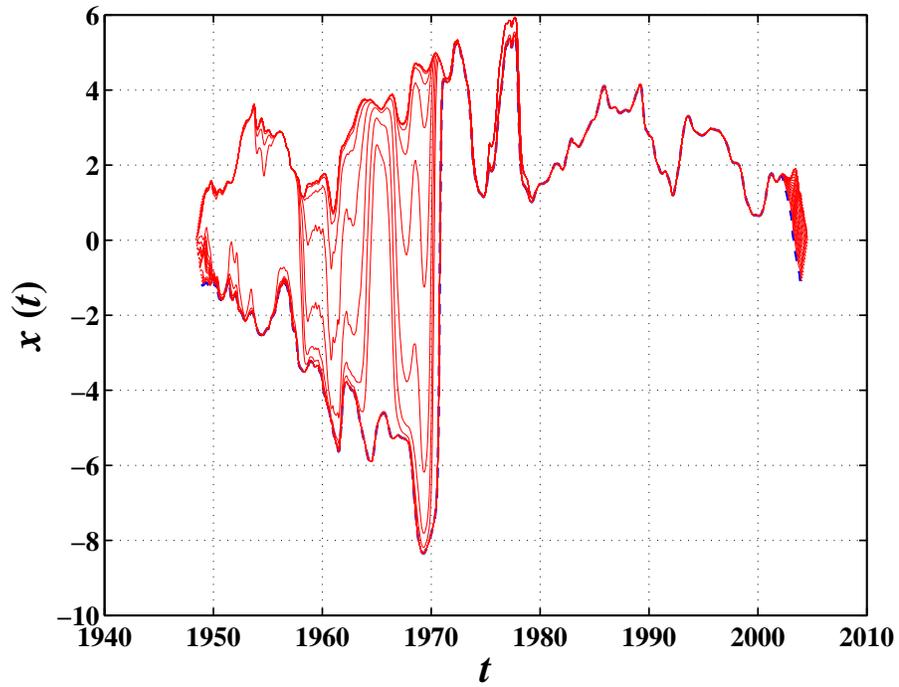}
\caption{Same as figure \ref{FigSZC_InfGDP5Tau} for the pair
(inflation, unemployment rate). Positive values
of $\tau(t)=x(t)$ correspond
to the unemployment lagging behind or being caused by inflation.
The dashed blue line is the optimal path with the minimal ``energy.'' }
\label{FigSZC_InfUnemp5Tau}
\end{figure}

\end{document}